\documentclass[useAMS,usenatbib]{mn2e}
\title[Precursors in Pulsar Profiles]
{On the Nature of Precursors in the Radio Pulsar Profiles}

\author[S. A. Petrova]{S. A. Petrova
\thanks{E-mail: petrova@ira.kharkov.ua}\\
Institute of Radio Astronomy, NAS of Ukraine, 4, Chervonopraporna
Str., 61002 Kharkov, Ukraine}

\begin{document}
\date{Received\dots}

\pagerange{\pageref{firstpage}--\pageref{lastpage}} \pubyear{2007}

\maketitle

\label{firstpage}

\begin{abstract}
In the average profiles of several radio pulsars, the main pulse
is accompanied by the preceding component. This so called
precursor is known for its distinctive polarization, spectral, and
fluctuation properties. Recent single-pulse observations hint that
the sporadic activity at the extreme leading edge of the pulse may
be prevalent in pulsars. We for the first time propose a physical
mechanism of this phenomenon. It is based on the induced
scattering of the main pulse radiation into the background. We
show that the scattered component is directed approximately along
the ambient magnetic field and, because of rotational aberration
in the scattering region, appears in the pulse profile as a
precursor to the main pulse. Our model naturally explains high
linear polarization of the precursor emission, its spectral and
fluctuation peculiarities as well as suggests a specific
connection between the precursor and the main pulse at widely
spaced frequencies. This is believed to stimulate multifrequency
single-pulse studies of intensity modulation in different pulsars.
\end{abstract}

\section{Introduction}

The radio profile of a pulsar may include the precursor - a
peculiar component preceding the main pulse by up to a few tens
degrees in longitude. The presence of such a component is more or
less firmly ascertained for the Crab pulsar \citep*{c70}, PSR
B1055-52 \citep{m76}, PSR B1822-09 \citep*{f81}, and the Vela
pulsar \citep{kd83}. The precursor differs substantially from the
components of the main pulse by its spectral and polarization
properties. In the main pulse, the component widths and
separations generally increase with wavelength, signifying the
overall broadening of the emission cone. At the same time, the
width of the precursor and its separation from the main pulse
remains unchanged over a broad frequency range. The spectrum of
the precursor is also distinct. In the pulsar B1822-09, it is
unusually flat \citep{f81,g94}, whereas in the Crab pulsar it is
extremely steep \citep{m71}. The most distinctive feature of the
precursor is its complete linear polarization, in contrast to the
main pulse emission, which is typically somewhat depolarized due
to simultaneous presence of the two orthogonally polarized modes.

The single-pulse studies have further revealed some fascinating
features of the precursor emission. The precursor intensity shows
strong pulse-to-pulse fluctuations. In the Vela pulsar, the
intensity variations affect the longitudinal location of the
precursor: the stronger the precursor, the larger is its
separation from the main pulse \citep{kd83}. In PSR B1822-09, the
precursor appears pronounced only in strong enough pulses; weak
pulses, instead, include the interpulse component, which lags the
main pulse by about $180^\circ$ in longitude \citep{f81,f82,g94}.

Because of sporadic character of the precursor emission, it does
not necessarily form a component in the average profile. For
example, the pulsars B0950+08 and B1656+14 do not exhibit a
precursor in the average profile, but occasional pulses do
incorporate strong subpulses at the extreme leading edge of the
pulse \citep{hc81,welt06}. Thus, in a broad sense, the precursor
phenomenon can be defined as a sporadic activity in the
longitudinal region preceding the main pulse. One can expect that
the precursors defined in this way are much more typical of
pulsars than is usually believed.

It is interesting to note that the precursor phenomenon appears
related to other manifestations of intensity modulation in pulsar
radio emission. Recent single-pulse observations of PSR J1326-6700
\citep*{w07} reveal the precursor component arising exceptionally
during the nulls of the main pulse, when the intensity of the
latter drops below the detection level. Furthermore, several other
pulsars of the above survey exhibit occasional alternation between
the two forms of the average profile (mode changing phenomenon),
the modal profiles being mainly different in the intensity of the
leading component relatively to that of the rest of the profile.

Unusual spectral, polarization, and fluctuation properties of the
precursors make them puzzling. To the best of our knowledge, no
attempts have previously been aimed at explaining the physics of
this phenomenon. Recently \citet*{d05} have suggested a
geometrical model for the profile of PSR B1822-09. These authors
assume that the main pulse and the precursor originate
independently at different locations in the magnetosphere and the
precursor emission intermittently reverses its direction to form
the interpulse. In that model, the mechanism of reversal of the
emission direction remains obscure, but for any conceivable
switching mechanism it is principally difficult to explain its
dependence on the main pulse intensity.

In the present paper, we for the first time propose a physical
mechanism of the precursor formation. In our model, the precursor
arises as a result of induced scattering of the main pulse
emission into the background by the particles of the
ultrarelativistic highly magnetized plasma of a pulsar. Our
mechanism naturally explains the observed polarization, spectral,
and fluctuation properties of the precursor emission as well as
suggests its connection to the main pulse.

\section{MAGNETIZED INDUCED SCATTERING}
\subsection{Statement of the problem}
Pulsar radio emission is generated deep in the magnetosphere
inside of the open field line tube. Hence, it originates and
propagates in the flow of the ultrarelativistic electron-positron
plasma, which streams along the open magnetic lines. As the
brightness temperatures of the radio emission are extremely high,
$T_B\sim 10^{25}-10^{30}$ K, the waves may be subject to efficient
induced scattering off the plasma particles. According to the
radio emission theories based on the plasma instabilities, the
frequency of the generated waves is close to the local
Lorentz-shifted proper plasma frequency,
$\omega\sim\omega_p\sqrt{\gamma}$, where $\omega_p\equiv\sqrt{4\pi
n_ee^2/m}$, $n_e$ is the plasma number density, $e$ and $m$ are
the electron charge and mass, and $\gamma$ is the plasma
Lorentz-factor \citep[but see][for the criticism of this
point]{gm99}. In the vicinity of the emission region, where the
above condition is still valid, induced scattering off the plasma
particles is a collective process. The transverse waves are
involved in the induced three-wave interactions \citep{lm06}. A
particular case of induced Raman scattering in application to the
pulsar magnetosphere has been considered by \citet{gk93} and
\citet{l98}.

As the plasma number density decreases with distance from the
neutron star along with the magnetic field strength, $n_e\propto
B\propto r^{-3}$, well above the emission region
$\omega\gg\omega_p\sqrt{\gamma}$ and the collective effects become
negligible. The external magnetic field significantly affects the
scattering process on condition that the radio wave frequency in
the particle rest frame is much less than the electron
gyrofrequency, $\omega^\prime\ll\omega_G\equiv eB/mc$. This
condition is valid up to the radius of cyclotron resonance, which
lies in the outer magnetosphere. In the present paper, we examine
the induced scattering, which takes place well above the emission
region and well below the cyclotron resonance radius. Then the
magnetized induced Compton scattering is a single-particle process
and the incident waves are approximately transverse
electromagnetic waves polarized either in the plane of the
wavevector and the ambient magnetic field or perpendicularly to
this plane.

In application to pulsar magnetosphere, induced scattering in a
superstrong magnetic field has first been considered by
\citet{bs76} and found to be efficient. Later on the process has
been suggested to explain a number of phenomena in pulsar radio
emission \citep{lp96,p04a,p04b}. In the present paper, we consider
the pulsar radio beam scattering into the background and
particularly concentrate on the growth of the scattered component,
which is identified with the precursor component of the pulse
profile.

\subsection{Basic equations}
In the preceding literature on the induced scattering in a
superstrong magnetic field, the kinetic equation for photon
occupation numbers is derived from an analysis of the scattering
by a single relativistic electron and does not include the
particle distribution function explicitly, so that it directly
corresponds to the cold plasma case. In the present paper, we
generalize the kinetic equation for the more realistic case of a
hot plasma. The corresponding formalism has also been outlined in
\citet{bs76}, but we shall go through the derivation once more in
order to correct a slight error in that paper.

In the approximation of an infinitely strong magnetic field, the
scattering cross-section in the electron frame takes the form
\begin{equation}
\frac{{\rm d}\sigma}{{\rm
d}\Omega_1^\prime}=r_e^2\sin^2\theta^\prime\sin^2\theta_1^\prime\,,
\end{equation}
where $r_e$ is the classical electron radius, $\theta^\prime$ and
$\theta_1^\prime$ are the propagation angles of the incident and
scattered photons, respectively, $\rm d\Omega_1^\prime$ is the
elementary solid angle for the scattered photons, and the primes
denote the quantities of the electron rest frame. The
cross-section (1) corresponds to the scattering between the
photons with the ordinary polarization, i.e with the electric
vectors lying in the plane of the ambient magnetic field. The
scattering involving the extraordinary polarization states is
negligible, since any perturbed motion of a particle (in the field
of the incident wave) perpendicular to the ambient magnetic field
is suppressed. Let us consider the scattering in the laboratory
frame between the two photon states, $\bmath{k}$ and $\bmath
{k_1}$, involving the electrons with the momenta $p$ and $p+\delta
p$ along the magnetic field. In the scattering act, the momentum
parallel to the magnetic field is conserved:
\begin{equation}
\delta p=\hbar k\cos\theta-\hbar k_1\cos\theta_1\,.
\end{equation}
The probability of generating spontaneously scattered photons per
electron per unit time is given by
\begin{equation}
\frac{{\rm d}P}{{\rm d}t}=n({\bmath k})\frac{{\rm d}^3{\bmath
k}}{(2\pi)^3}\eta\frac{{\rm d}\sigma }{{\rm
d}\Omega_1}c^4\frac{\rm d^3\bmath{k_1}}{\omega_1^2}\delta\left
(\omega_1-\frac{\eta}{\eta_1}\omega\right )\,,
\end{equation}
where $n(\bmath k)$ is the photon occupation number, $\eta\equiv
1-\beta\cos\theta$, $\beta$ is the particle velocity in units of
$c$, and the argument of the delta-function signifies the equality
of the initial and final frequencies in the particle rest frame:
$\omega\gamma\eta=\omega_1\gamma\eta_1$. The cross-section in the
laboratory frame is related to equation (1) by means of
relativistic transformations, ${\rm d}\Omega_1^\prime=\rm
d\Omega_1/(\gamma\eta_1)^2$, $\sin\theta^\prime =\sin\theta
/(\gamma\eta)$. The rate of change of the photon occupation number
as a result of induced scatterings is
\begin{equation}
\frac{{\rm d}n}{{\rm d}t}\frac{{\rm d}^3\bmath k}{(2\pi)^3}=\int
\left [f(p+\delta p)-f(p)\right ]\frac{{\rm d} P}{{\rm d}t}n_1{\rm
d}p\,,
\end{equation}
where $f(p)$ is the electron distribution function. Note that
\citet{bs76} write down literally the same formula, which
incorporates the distribution function in Lorentz-factor,
$N(\gamma)$ (cf. eq.[21] in that paper), but it is not true. As
$p\equiv\beta\gamma mc$, then $N(\gamma)=f(p){\rm d}p/{\rm
d}\gamma=mcf(p)/\beta$, where $\beta$ is also an implicit function
of $p$, i.e. $\beta(p+\delta p)\neq\beta(p)$. In the above kinetic
equation, ${\rm d}/{\rm d}t\equiv \partial/\partial
t+c\partial/\partial r$. Hereafter we consider the stationary
case, omitting the explicit time derivative. Taking into account
that $f(p+\delta p)-f(p)\approx \delta p(\partial f/\partial p)$
and integrating equation (4) by parts yield
\begin{equation}
c\frac{\partial n}{\partial r}\frac{{\rm d}^3{\bmath
k}}{(2\pi)^3}=-\int f(p)\frac{\partial}{\partial p}\left [\delta
p\frac{{\rm d}P}{{\rm d}t}n_1 \right]{\rm d}p\,.
\end{equation}
This can be easily rewritten in terms of $N(\gamma)$, and
incorporating equations (1)-(3) we obtain
\begin{eqnarray}
\frac{\partial n}{\partial r}=\int {\rm d}\gamma
N(\gamma)\int\frac{{\rm d}\Omega_1r_e^2c^4\hbar \beta
(\cos\theta_1-\cos\theta)}{mc^3}\nonumber \\
\times\frac{\partial}{\partial \gamma}\left
[\frac{nn_1\sin^2\theta\sin^2\theta_1\delta
(\omega_1\eta_1-\omega\eta)k_1^2{\rm
d}k_1}{\gamma^6\eta^2\eta_1^3\omega_1}\right ].
\end{eqnarray}
The difference of the above equation from eq.[23] in \citet{bs76}
lies in that the factor $\beta$ is not subject to differentiation
with respect to $\gamma$. Integration of equation (6) with the
help of delta-function and subsequent differentiation yield
finally
\begin{eqnarray} \frac{\partial n}{\partial r}=\frac{\hbar
r_e^2n\sin^2\theta}{mc}\int N{\rm
d}\gamma\int\left\{\frac{(\eta_1-\eta)^2}
{\eta_1^2\gamma^3}\frac{\partial k_1^2n_1}{\partial k_1}
\right.\nonumber \\ \left. +\frac{6k\eta^2(\eta_1-\eta)n_1}
{\gamma\eta_1^2}\left
[1-\frac{\eta_1+\eta}{2\gamma^2\eta\eta_1}\right ] \right\}\frac
{\sin^2\theta_1{\rm d}\Omega_1} {\gamma^6\eta^3\eta_1^3\beta^2}.
\end{eqnarray}
For the monoenergetic distribution of the particles, this exactly
coincides with the kinetic equation obtained by \citet{bs76} in
another way (cf. their eq. [30]). Our equation (7) presents a
correct generalization for an arbitrary distribution function of
particles. One can see that in general the detailed form of the
particle distribution does not play a crucial role, and hence the
characteristic features of the scattering studied below are
determined by the role of the external magnetic field. Thus, the
evolution of the photon occupation numbers may well be described
by incorporating the monoenergetic distribution with some
characteristic Lorentz-factor. Seeing that the actual distribution
function of the plasma in pulsar magnetosphere is still obscure,
this is really a proper assumption.

\section{RADIO BEAM SCATTERING INTO BACKGROUND}
We are interested in the problem of induced scattering of pulsar
radio beam into the background. Far from the emission region, the
pulsar radiation propagates quasi-transversely with respect to the
ambient magnetic field, $1/\gamma\ll\theta_{\rm b}<1$, and it is
known to be highly directional, i.e. $\Delta\theta_{\rm b}\la
1/\gamma$. Therefore at any point of the scattering region the
incident radiation can be represented by a single wavevector
${\bmath k}_{\bf b}$. In case of efficient scattering, the
orientations of the scattered photons are believed to concentrate
near $\theta_{\rm bg}^{\rm max}$, which corresponds to the maximum
scattering probability at a fixed $\theta_{\rm b}$. (Note that the
kinetic equation (7) is independent of the azimuthal coordinates
of the photons.)

To specify $\theta_{\rm bg}^{\rm max}$ we investigate the
evolution of the background occupation numbers using equation (7)
and prescribing $n(k,\theta)$ to the background and
$n_1(k_1,\theta_1)$ to the beam. Given that $\theta_{\rm b}\gg
1/\gamma$ the second term on the right-hand side is
$\theta_b^2\gamma^2$ times larger than the first one and decreases
with $\theta_{\rm b}$ as $\vert\theta_{\rm b}^2-\theta_{\rm
bg}^2\vert/\theta_{\rm b}^4$. Therefore one can expect that the
whole integrand peaks at $\theta_{\rm bg}^{\rm max}\sim 1/\gamma$.
Exact numerical calculations of the scattering probability in the
range $\theta_{\rm bg}\la 1/\gamma$ reveal the following features.
The rate of increase of the background occupation numbers does
peak at $\theta_{\rm bg}^{\rm max}\sim 1/\gamma$, the peak width
at half maximum is $\la 1/\gamma$, and the first term of equation
(7) can still be ignored. In the kinetic equation describing the
rate of change of the beam occupation numbers (which is given by
equation (7) with $\theta=\theta_{\rm b}$ and
$\theta_1=\theta_{\rm bg}^{\rm max}$), the first term is also
negligible as compared to the second one.

As $1/\gamma$ is a small parameter, we examine induced scattering
between the two photon states with the fixed polar angles
$\theta\gg 1/\gamma$ and $\theta_1=\theta_{\rm bg}^{\rm max}\sim
1/\gamma$ and the frequencies related as $\omega_1=\omega\eta
/\eta_1\sim\omega\theta^2\gamma^2\gg\omega$. It is convenient to
introduce the photon intensities,
$i_{\nu_a}=\hbar\omega^3n({\bmath k})/2\pi^2c^2$ and
$i_{\nu_b}=\hbar\omega_1^3n_1({\bmath k_1})/2\pi^2c^2$, where
$\nu_a\equiv\omega/2\pi$ and $\nu_b\equiv\omega_1/2\pi$. The
angular distributions have the form of delta-function and one can
integrate them over the solid angle to obtain the spectral
intensities $I_{\nu_{a,b}}=\int i_{\nu_{a,b}}{\rm d}\Omega_{a,b}$.
Then the evolution of the beam intensities in the course of
induced scattering is approximately described as \citep[see
also][]{p04b}:
\begin{eqnarray}
\frac{{\rm d}I_{\nu_a}}{{\rm d}r}=-aI_{\nu_a}I_{\nu_b}\,,\nonumber
\\ \frac{{\rm d}I_{\nu_b}}{{\rm d}r}=aI_{\nu_a}I_{\nu_b}\,,
\end{eqnarray}
where
\begin{equation}
a\sim\frac{24n_er_e^2}{m\nu_a^2\gamma^5\theta^4}
\end{equation}
and $n_e$ is the particle number density, $n_e\equiv\int
N(\gamma){\rm d}\gamma$. The above system of equations has the
first integral,
\begin{equation}
I_{\nu_a}+I_{\nu_b}={\rm const}\equiv I\,.
\end{equation}
Thus, our approximate consideration leads to the conservation of
the total intensity of the two beams. This actually implies that
the intensity redistribution between the beams is much more rapid
than the change of the total intensity $I$, and below we
concentrate on the photon transfer between the beams. The photons
are mainly transferred to the higher-frequency state,
$\nu_b\gg\nu_a$, and become almost aligned with the ambient
magnetic field, $\theta_1\sim 1/\gamma$. This contrasts with the
non-magnetic scattering, in which case the kinetic equation
contains only the term qualitatively similar to the first term in
equation (7), the maximum scattering rate is in the direction
antiparallel to the particle velocity, and the photons shift
monotonically toward lower frequencies.

The solution of the system (8) is given by
\begin{eqnarray}
I_{\nu_a}=\frac{I(I_{\nu_a}^{(0)}/I_{\nu_b}^{(0)})
\exp(-Iar)}{1+(I_{\nu_a}^{(0)}/I_{\nu_b}^{(0)})\exp(-Iar)}\,,\nonumber
\\ I_{\nu_b}=\frac{I}{1+(I_{\nu_a}^{(0)}/I_{\nu_b}^{(0)})\exp(-Iar)}\,.
\end{eqnarray}
One can see that the efficiency of intensity transfer between the
beams is determined by $\Gamma\equiv Iar$. At large enough
$\Gamma$, $I_{\nu_a}\to 0$ and $I_{\nu_b}\to I$. Since the initial
intensity of pulsar beam greatly exceeds that of the background
radiation, $I_{\nu_b}^{(0)}/I_{\nu_a}^{(0)}\ll 1$, the intensity
transfer to the background is significant on condition that
\begin{equation}
\left (I_{\nu_b}^{(0)}/I_{\nu_a}^{(0)}\right )\exp(\Gamma)\ga 1\,.
\end{equation}
Note that $\Gamma$ includes the total intensity $I$, i.e. actually
$I_{\nu_a}^{(0)}$, and for large enough $\Gamma$ the scattering is
efficient independently of the smallness of the background
intensity.

\section{NUMERICAL ESTIMATES}
Now we are to  estimate the scattering efficiency, $\Gamma =Iar$,
where $I\approx I_{\nu_a}^{(0)}$ and $a$ is given by equation (9),
for the parameters relevant to pulsar magnetosphere. The spectrum
of pulsar radiation generally has the power-law form,
$I_{\nu_a}^{(0)}=I_{\nu_0}(\nu_a/\nu_0)^{-\alpha}$, and the
spectral intensity at frequencies $\nu_0\sim 100$ MHz is related
to the total radio luminosity of a pulsar, $L$, as
$I_{\nu_0}=L/\nu_0S$, where $S=\pi r^2w^2/4$ is the cross-section
of the pulsar beam at a distance $r$, $w$ the pulse width in the
angular measure. It is convenient to normalize the number density
of the scattering particles to the Goldreich-Julian number
density, $n_e=\kappa B/Pce$, where $\kappa$ is the plasma
multiplicity factor, and $P$ is the pulsar period. Keeping in mind
that $n_e\propto B\propto r^{-3}$, one can estimate the scattering
efficiency as
\begin{eqnarray}
\Gamma=10P^{-1}\frac{L}{10^{28}\,{\rm erg\,
s^{-1}}}\frac{B_\star}{10^{12}\,{\rm G}}\left
(\frac{\nu_a}{10^{8}\,{\rm Hz}}\right )^{-2-\alpha}\nonumber
\\ \times\frac{\kappa}{10^2}\frac{10^2}{\gamma}\left (\frac{w}{0.4}\right )^{-2}\left
(\frac{\theta}{0.1}\frac{\gamma}{10^2}\frac{r}{10^{8}\,{\rm
cm}}\right )^{-4},
\end{eqnarray}
where $B_\star$ is the magnetic field strength at the stellar
surface, and it is taken that the neutron star radius is $10^6$
cm. All the quantities in equation (13) are normalized to their
typical values. One can see that the scattering efficiency may be
as large as about a few tens. To conclude whether the condition of
efficient intensity transfer given by equation (12) can indeed be
satisfied let us estimate the level of background radiation
resulting from the spontaneous scattering of the pulsar radio
beam, $I_{\nu_b}^{(0)}\sim I_{\nu_a}^{(0)}n_e\eta r{\rm
d}\sigma/{\rm d}\Omega_1$. Taking into account that ${\rm d}\sigma
/{\rm
d}\Omega_1=r_e^2\sin^2\theta\sin^2\theta_1/(\gamma^6\eta_1^4\eta^2)$,
$\eta\approx \theta^2/2$ and $\eta_1\approx 1/\gamma^2$, we find
that
\begin{equation}
\frac{I_{\nu_b}^{(0)}}{I_{\nu_a}^{(0)}}\sim 2n_er_e^2r\sim
10^{-10}P^{-1}\frac{\kappa}{10^2}\frac{B_\star}{10^{12}\,{\rm
G}}\left (\frac{r}{10^8\,{\rm cm}}\right )^{-2}.
\end{equation}
Hence, typically $I_{\nu_b}^{(0)}/I_{\nu_a}^{(0)}\sim
10^{-8}-10^{-12}$, and to satisfy equation (12) $\Gamma=18-28$ are
necessary. Thus, the intensity transfer from the radio beam to the
background can indeed be significant in pulsars, especially at low
enough frequencies.

Let us consider the location of the scattering region in the
magnetosphere of a pulsar. According to equation (13), $\Gamma$
shows strong explicit dependence on $r$, $\Gamma\propto r^{-4}$.
However, the implicit dependence on $r$ appears still more
significant. As $\theta\propto r$ (see below),
$\nu_a=\nu_b/\theta^2\gamma^2\propto r^{-2}$. This can be
understood as follows. At a fixed frequency $\nu_b$, the
background radiation actually grows because of the scattering of
different radio beam frequencies $\nu_a$, which satisfy the
condition $\nu_b=\nu_a\theta^2(r)\gamma^2$ at different altitudes
$r$. Larger altitudes imply lower frequencies $\nu_a$, in which
case the incident intensity of the radio beam is much larger and
stimulates stronger scattering. Taking into account the above
considerations, one can obtain that $\Gamma\propto r^{2\alpha
-4}$. Hence, the scattering efficiency increases with distance at
$\alpha>2$ and decreases at $\alpha<2$. At high enough
frequencies, $\ga 1$ GHz, three of the pulsars with the precursor
components, namely the Crab, the Vela and PSR B1822-09, have
$\alpha=2.8$, 2.7 and 2.3, respectively, whereas for PSR B1055-52
there is no spectral data in this region (perhaps, the spectrum is
too steep for the pulsar to be detectable). At lower frequencies
such a steep spectrum is preserved only in the Crab pulsar, while
in the other pulsars under consideration $\alpha$ drops below 2.
Note, however, that at lower frequencies the scattering is more
efficient and may noticeably suppress the radio beam intensity,
leading to the pulsar spectrum flattening, so that the original
spectra may be much steeper than the observed ones. Thus, we
assume that $\Gamma$ increases with altitude and, correspondingly,
the induced scattering is most efficient at the upper boundary of
the scattering region, i.e. at distances of order of the cyclotron
resonance radius, which is defined as
$2\pi\nu_a\gamma\theta^2/2=\omega_G$ and estimated as
\begin{equation}
\frac{r_c}{r_L}=0.4\left (\frac{B_\star}{10^{12}\,{\rm
G}}\frac{10^2}{\gamma}\frac{10^8\,{\rm Hz}}{\nu_a}\right
)^{1/5}\left (\frac{1\,{\rm s}}{P}\right )^{3/5},
\end{equation}
where $r_L$ is the light cylinder radius and it is taken that
$\theta=r/2r_L$ (see below). One can see that the region of
cyclotron resonance typically lies in the outer magnetosphere, at
distances of order of the light cylinder radius.

The location of the scattered component in the pulse profile can
be examined as follows. Since the scattering region lies well
above the emission region, the wavevector ${\bmath k}$ makes the
angle $\sim r\sin\zeta /r_L$ with the instantaneous direction of
the magnetic axis of the rotating magnetosphere (here $\zeta$ is
the angle between the rotational and magnetic axes of a pulsar).
As the polar angle of the point of scattering is $r\sin\zeta
/r_L$, in the dipolar geometry the local magnetic field vector
${\bmath b}$ is inclined at the angle $3r\sin\zeta /2r_L$ to the
magnetic axis, so that the angle between ${\bmath k}$ and ${\bmath
b}$ is $r\sin\zeta /2r_L$. In the corotating frame, the scattered
radiation is directed approximately along ${\bmath b}$. Then in
the laboratory frame it is shifted because of rotational
aberration by the angle $r\sin\zeta /r_L$ in the direction of
rotation, i.e. toward the magnetic axis. Thus, the scattered
component precedes the main pulse by $\Delta\lambda\sim r\sin\zeta
/2r_L$ in longitude. One can see that the precursor separation
from the main pulse is determined by the height of the scattering
region, $r$, and does not exceed $\sim 30^\circ$ for the
scattering inside the light cylinder.

\section{DISCUSSION}
We have considered the induced Compton scattering by the particles
of the ultrarelativistic electron-positron plasma in the presence
of a  superstrong magnetic field. In particular, we have examined
the scattering of pulsar radio beam into background, which takes
place in the open field line tube of a pulsar. It has been
demonstrated that the photons are predominantly scattered
approximately along the ambient magnetic field. This contrasts
with the non-magnetic scattering, in which case the scattered
photons concentrate in the backward direction. This difference is
solely determined by a specific role of the superstrong magnetic
field in the scattering process and does not depend on a detailed
form of the particle distribution function.

Induced scattering in a superstrong magnetic field transfers the
photons from lower to higher frequencies,
$\nu_b\sim\nu_a\theta^2\gamma^2\sim n\cdot 10\nu_a$, and if the
process is efficient, the scattered component may become as strong
as the original radio beam, $I_{\nu_b}(\nu_b)\sim
I_{\nu_a}^{(0)}(\nu_a)$. As the beam has a decreasing spectrum,
$I_{\nu_a}^{(0)}(\nu_a)\gg I_{\nu_a}^{(0)}(\nu_b)$, the intensity
of the scattered component may dominate the original beam
intensity at the same frequency $\nu_b$.

For steep enough original spectra of pulsar radiation, $\alpha>2$,
the induced scattering in a superstrong magnetic field is most
efficient at distances roughly comparable to the radius of
cyclotron resonance. Because of rotational aberration, the
scattered component appears in the pulse profile as a precursor to
the main pulse. This effect provides the main pulse-precursor
separations in longitude $\Delta\lambda\sim r\sin\zeta/2r_L$,
which may run up to $\sim 30^\circ$. Since the length of the
scattering region is larger than the height of the emission
region, the intrinsic radius-to-frequency mapping of the radio
emission is smeared. The effective height of the scattering region
is an extremely weak function of the wave frequency, so that the
main pulse-precursor separation is practically independent of
frequency, just as is observed.

Since the induced scattering in the superstrong magnetic field
holds only between the ordinary waves, the scattered component
should have complete linear polarization. This is indeed the main
distinctive feature of the observed precursors. Note that in
general $[{\bmath k}\times{\bmath b}]\not\parallel[{\bmath
k_1}\times{\bmath b}]$, i.e. in the main pulse and precursor the
position angles of linear polarization should somewhat differ.
Such a difference can be noticed, e.g., in PSR B1822-09
\citep{f81}. Besides that, if the main pulse is dominated by the
extraordinary rather than ordinary polarization, the position
angle of the precursor should additionally differ by $90^\circ$,
as is the case in the Vela pulsar \citep{kd83}.

As first noted by \citet{f81}, the precursor components are met in
pulsars with relatively large surface magnetic field. Firstly,
large $B_\star$ are necessary for the regime of superstrong
magnetic field to hold well above the emission region. Secondly,
the scattering efficiency is proportional to $B_\star$. Short
periods and large radio luminosities also favour significant
scattering.

The pulse-to-pulse variations of the incident intensity and of the
physical parameters in the scattering region may result in strong
fluctuations of the precursor emission. The former variations
imply the main pulse-precursor connection, which may have
diversiform observational manifestations. For example, PSR
J1326-6700 shows occasional main pulse nullings accompanied by the
strong precursor emission \citep{w07}. This can be interpreted as
a consequence of extremely strong scattering,
$(I_{\nu_b}^{(0)}/I_{\nu_a}^{(0)})\exp (\Gamma)\gg 1$, when the
main pulse intensity is almost completely transferred to the
precursor, $I_{\nu_a}\to 0$, $I_{\nu_b}\to I_{\nu_a}^{(0)}$. Note
that this may happen only if the original intensity is mainly in
the ordinary mode, which is subject to the scattering. In case of
a moderately strong scattering,
$(I_{\nu_b}^{(0)}/I_{\nu_a}^{(0)})\exp (\Gamma)\sim 1$, the main
pulse intensity $I_{\nu_a}^{(0)}$ is almost unchanged, whereas the
precursor grows exponentially with $I_{\nu_a}^{(0)}$. Therefore
even weak fluctuations of the latter quantity may affect the
scattered component dramatically. In some pulsars the precursors
are indeed met only in strong pulses \citep{hc81,g94,welt06}, and
one can expect that the transient precursors are much more
abundant in the pulsar population and are yet to be studied
observationally.

The precursor component can fluctuate not only in intensity but
also in pulse longitude. In the Vela pulsar, stronger precursors
exhibit larger separations from the main pulse, which is thought
to result from the fluctuations of the physical parameters in the
scattering region. Larger separations imply larger scattering
heights, $\Delta\lambda\propto r$, in which case the angle of
incidence of the photons is also larger, $\theta\propto r$, and at
a fixed frequency $\nu_b$ the precursor is formed by the photons
coming from lower frequencies $\nu_a=\nu_b/\theta^2(r)\gamma^2$,
which are more numerous and stimulate stronger scattering.

\bsp

\label{lastpage}

\end{document}